%% file: paper.tex
 \documentclass[sigconf,10pt]{acmart}

\newcommand{\floor}[1]{\lfloor {#1} \rfloor}

\usepackage{subcaption}
\usepackage{enumitem}
\usepackage{amsfonts,amsmath,amsthm}

\AtBeginDocument{%
  }

\newcommand{\xref}[1]{\S\ref{#1}}

\newcommand{\squishlist}{\begin{itemize}[itemsep=1pt,parsep=2pt,topsep=3pt,partopsep=0pt,leftmargin=0em, itemindent=1em,labelwidth=1em,labelsep=0.5em]}
\newcommand{\squishend}{\end{itemize}}



\copyrightyear{2025}
\acmYear{2025}
\setcopyright{cc}
\setcctype{by}
\acmConference[ACM MOBICOM '25]{The 31st Annual International Conference
on Mobile Computing and Networking}{November 4--8, 2025}{Hong Kong, China}
\acmBooktitle{The 31st Annual International Conference on Mobile Computing
and Networking (ACM MOBICOM '25), November 4--8, 2025, Hong Kong,
China}\acmDOI{10.1145/3680207.3765251}
\acmISBN{979-8-4007-1129-9/2025/11}
\begin{document}




\title{Wireless Hearables With  Programmable Speech AI Accelerators}

 \author{Malek Itani}
 \affiliation{Paul G. Allen School,  University\\ of Washington, Seattle, WA  
 \country{USA}
 }
 \email{malek@cs.washington.edu }

 \author{Tuochao Chen}
 \affiliation{Paul G. Allen School,  University\\ of Washington, Seattle, WA  
 \country{USA}
 }
 \email{tuochao@cs.washington.edu }

 \author{Arun Raghavan}
 \affiliation{Department of Otolaryngology,  University of Washington  
 \country{USA}
 }
 \email{amraghav@uw.edu}

 \author{Gavriel Kohlberg}
 \affiliation{Department of Otolaryngology,  University of Washington 
 \country{USA}
 }
 \email{kohlberg@uw.edu}

 \author{Shyamnath Gollakota}
 \affiliation{Paul G. Allen School,  University\\ of Washington, Seattle, WA  
 \country{USA}
 }
 \email{gshyam@cs.washington.edu }



\input{abstract-3}





\maketitle

\pagestyle{plain}

\input{intro-1}
\input{system-1}
\input{results-1}

\input{related-1}
\input{discuss-1}
\input{conclude-2}

\balance
\bibliographystyle{ACM-Reference-Format}
\bibliography{refs}

\end{document}

%% file: abstract-3.tex
\begin{abstract}

The conventional wisdom has been that designing ultra-compact, battery-constrained wireless hearables with on-device speech AI models is challenging due to the high computational demands of streaming deep learning models. Speech AI models require continuous, real-time audio processing, imposing strict computational and I/O constraints.

We present {\it NeuralAids,} a fully on-device speech AI system for wireless hearables, enabling real-time speech enhancement  and denoising on compact, battery-constrained devices. Our system bridges the gap between state-of-the-art deep learning for speech enhancement and low-power AI hardware by making three key technical contributions: 1) a wireless hearable platform integrating a speech AI accelerator for efficient on-device  streaming inference, 2) an optimized dual-path neural network designed for low-latency, high-quality speech enhancement, and 3) a hardware-software co-design that uses mixed-precision quantization and  quantization-aware training to achieve real-time performance under strict power constraints. Our system processes 6 ms audio chunks in real-time, achieving an inference time of 5.54 ms while consuming 71.6 mW.  In real-world evaluations, including a user study with 28 participants, our system outperforms prior on-device models in speech quality and  noise suppression, paving the way for next-generation intelligent wireless hearables that can enhance hearing entirely on-device.

\end{abstract}

%% file: intro-1.tex
\begin{figure}[t!]
\centering
\vskip -0.15in
\includegraphics[width=0.85\linewidth]{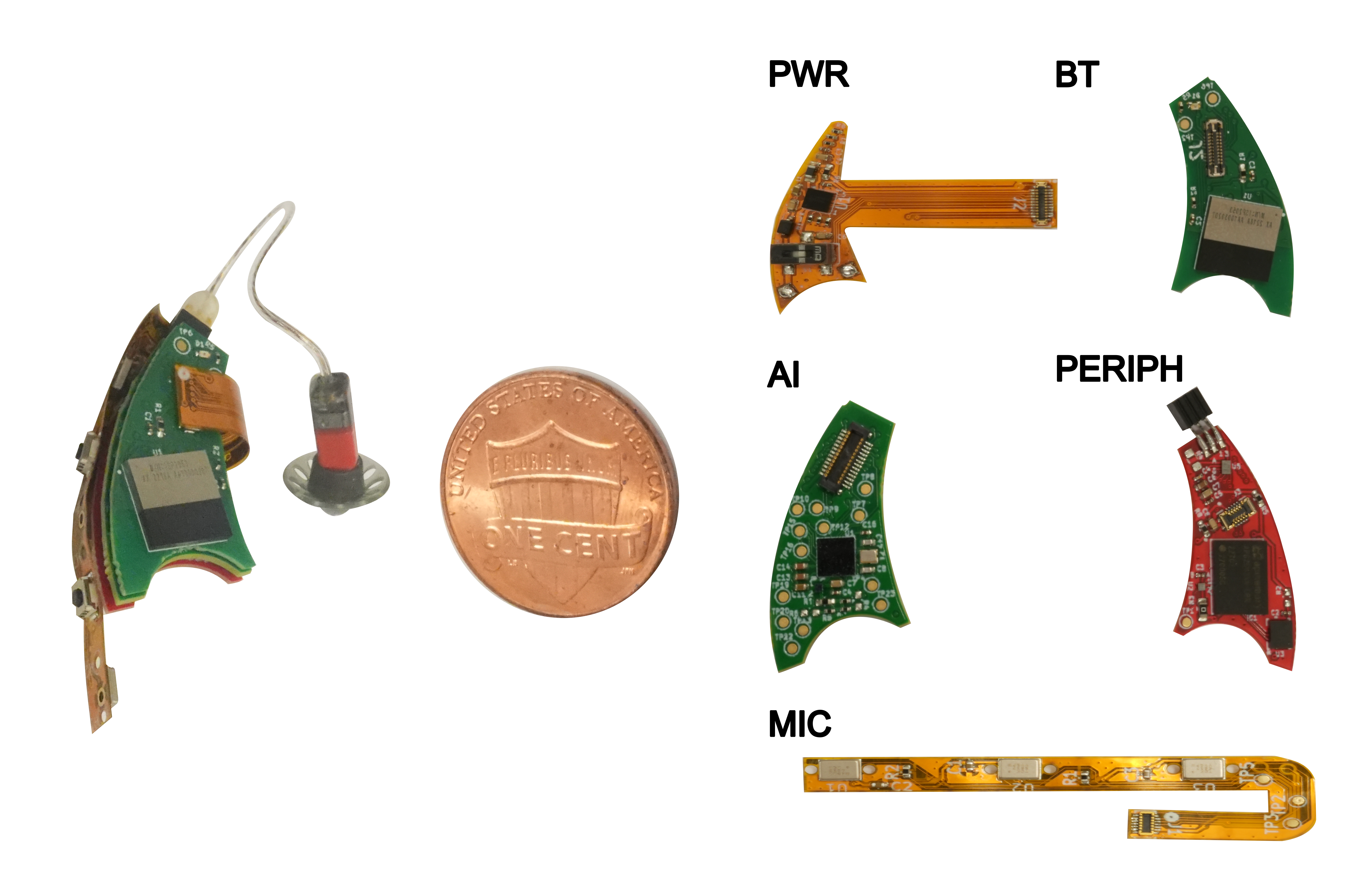}
\vskip -0.15in
\caption{NeuralAid hardware. \textmd{The device has five interconnected flexible and rigid circuit boards that together form an AI-enabled hearable. PWR (2-layer flexible PCB): manages power, charging, and programming, BT (2-layer rigid PCB): houses the BLE SoC,  AI (6-layer rigid PCB): contains a low-power AI accelerator for real-time speech AI, PERIPH (4-layer rigid PCB): hosts peripherals including RAM, NOR flash, IMU, and I2S DAC, MIC (2-layer flexible PCB): has a microphone array with three mics and two push buttons.}}
\label{fig:hardware}
\vskip -0.2in
\end{figure}

\section{Introduction}

In recent years, intelligent hearables have made significant advances in enhanced hearing, leveraging deep learning to  program acoustic scenes in real time~\cite{semantichearing,chatterjee2022clearbuds,soundbubble}. These  advancements enable  capabilities such as speech enhancement~\cite{10.1145/3539490.3539600}, noise suppression~\cite{chatterjee2022clearbuds}, and even target speech hearing~\cite{lookoncetohear}. However, current implementations rely on wired headsets and computationally demanding platforms like smartphones or high-power embedded systems~\cite{semantichearing,lookoncetohear,soundbubble}—devices with far greater processing power, memory, and energy budgets than what can realistically fit within ultra-compact, battery-constrained hearables like earbuds and  hearing aids.

The fundamental challenge is that streaming deep learning models are traditionally associated with high computational demands and power consumption. This raises key questions about whether real-time speech AI models can be deployed on small, battery-powered hearables. Offloading computations to a smartphone is unreliable due to the stringent sub-10 ms algorithmic latency requirements of enhanced and augmented audio applications, which are highly sensitive to wireless network variability, I/O delays, and operating system overhead on smartphones. Meanwhile, running speech AI models on-device has so far been constrained by the limited availability of AI accelerators that are both powerful enough for real-time speech AI and small enough to fit within these miniature wireless devices. Overcoming these barriers is essential to unlocking the next generation of truly intelligent and self-sufficient wireless hearables.


In this paper, we explore whether it is possible to design wireless hearables that can run real-time speech AI models entirely on-device while understanding the trade-offs between power consumption and efficiency. Achieving this requires addressing three key challenges.
\squishlist
    \item  Existing   research platforms for  hearables lack the computational capabilities required to run deep learning models for augmented and enhanced audio applications. In contrast to classification tasks, these models must continuously operate in a streaming manner, processing  audio input at a minimum sampling rate of 16~kHz. To do this, they should process audio in small chunks of around 6~ms,   maintain an inference time of under 6~ms to ensure real-time performance, and  continuously output audio at the same sampling rate.  Meeting these requirements imposes significant computational and I/O constraints. Further, due to the limited non-valatile storage available on these compact devices, the model size must not exceed 1.5 MB and to enable more than six hours of continuous operation on a 675 hearing aid battery, deep learning power consumption must be below 100~mW.

\begin{figure}[t!]
\centering
\includegraphics[width=0.55\linewidth]{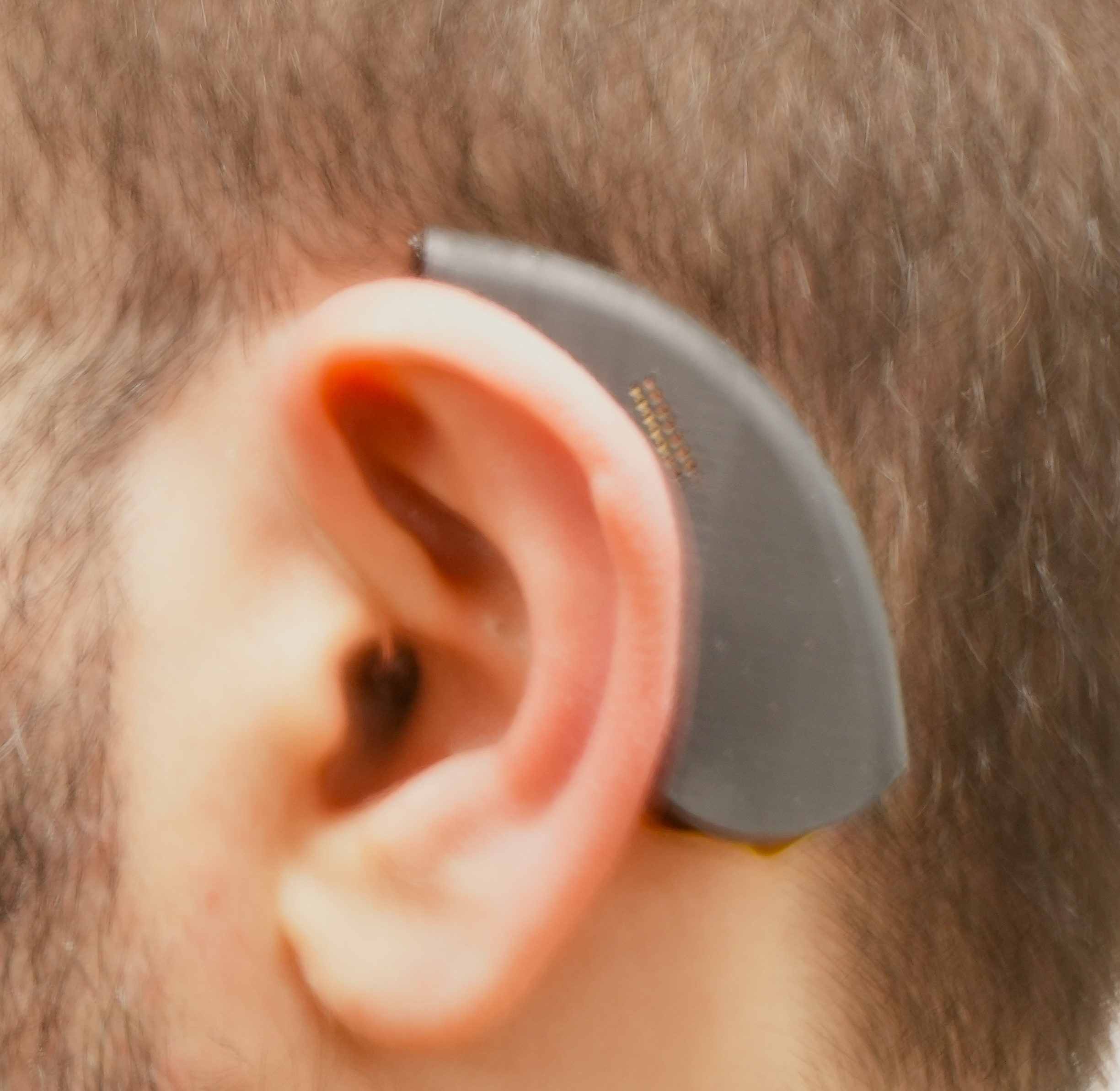}
\vskip -0.15in
\caption{\textmd{A person wearing a NeuralAid, which rests behind the ear, with a receiver inserted into their ear canal.}}
\label{fig:wear}
\vskip -0.05in
\end{figure}

    \item Tiny, low-power AI  accelerator hardware~\cite{syntiant,greenwaves} is  inherently far more constrained  than  GPUs and general-purpose embedded CPUs like the Raspberry Pi. On the other hand, effective neural architectures for speech enhancement rely on components such as convolutions~\cite{convtasnet}, LSTMs~\cite{dprnn}, transformers~\cite{sepformer}, and state-space layers~\cite{spmamba}.  These models prioritize speech quality over real-time, on-device, or low-power constraints --- indeed, transformers and state-space models exceed our target hardware’s runtime and memory limits. While efforts like TinyLSTM~\cite{tinylstms} and its successor  TinyDenoiser~\cite{tinydenoiser}   aim for on-device deployment, their 25 ms algorithmic latency falls short of the low-latency requirements for enhanced hearing applications. 

    \item While smartphones and general-purpose embedded devices  can run floating point operations, this increases the runtime and power consumption on  low-power hardware. In general, low-power AI accelerators function as a compute cluster with standard processors for general-purpose tasks and a dedicated neural accelerator optimized for low-power, highly quantized integer computations. However, quantizing all neural network weights, activations, and inputs can significantly degrade audio quality.  Further, different neural network layers are affected differently by quantization errors, necessitating mixed-precision quantization and compensation for non-linear errors at each layer. Thus,  achieving optimal performance requires a low-power hardware-software co-design approach.

\squishend

We introduce {\it NeuralAids}, which makes three key contributions across wireless hearable hardware, embedded systems, deep learning, and hardware-software co-design.

\squishlist
\item {\bf Wireless hearables with speech AI accelerators.} We design NeuralAids, a real-time speech AI system built on a modular hardware architecture with five stacked PCBs (Fig.~\ref{fig:hardware}). It integrates a low-power AI accelerator (GreenWaves GAP9) with an nRF53-based System-on-Chip (SoC) for Bluetooth Low Energy (BLE) connectivity, audio processing, sensors, and both volatile and non-volatile memory. The system  runs  speech AI models, using GAP9’s compute cluster for real-time neural network inference. Optimized power management, high-speed audio interfaces, and flexible storage enable continuous operation in a behind-the-ear  compact form factor (Fig.~\ref{fig:wear}), making NeuralAids a unique platform for real-time speech AI applications.


\item {\bf Real-time on-device efficient neural network.}  Prior on-device models like TinyDenoiser~\cite{tinydenoiser}  are {constrained} by their limited {complexity} and inability to achieve effective performance with sub-10ms latencies. Instead, we begin with dual-path models that process audio in the time-frequency (TF) domain, treating time and frequency components as sequences using recurrent networks~\cite{fspen,semamba}. These architectures achieve state-of-the-art performance by modeling both temporal and frequency relationships. However, they are computationally expensive, as running recurrent networks across each frequency prevents real-time operation. Thus, they cannot run in real time on our hardware. To address this, our proposed architecture compresses frequencies using a linear layer and replaces LSTMs with Gated Recurrent Units (GRUs). These architectural changes and others listed in~\xref{sec:nnchanges} achieve real-time performance by  reducing  runtime  while outperforming TinyDenoiser   for the  denoising task.

\item {\bf Hardware-software co-design.} 
Floating-point speech denoising  models cannot achieve real-time operation on our target  accelerator hardware. To address this, we explore various quantized model configurations and use quantization-aware training (QAT) to simulate non-linear quantization errors and fine-tune the network to mitigate error propagation across layers. This  narrows the performance gap between our mixed-precision  and fully floating-point speech  models.

\squishend

We evaluated our system for speech enhancement in noisy scenarios.
Our results are as follows:
\squishlist
\item Before quantization, our floating-point dual-path network architecture improves speech quality by 2.41~dB over TinyDenoiser. Applying quantization-aware training reduces the performance gap between our mixed-precision and floating-point networks from 7.86~dB to 0.57~dB. 
\item Our proposed mixed-precision network can process 6 ms audio chunks within 5.54~ms on the GAP9 processor, achieving real-time operation. Furthermore, the model uses only 299~kB of memory and consumes 71.64~mW on our hardware.
\item Our system generalizes to six real-world indoor and outdoor environments, as well as to wearer head motion. Our training relies solely on synthetic data and does not require any training data collection using our hearable hardware.
\item In a user study with 28 participants, our system achieved a higher mean opinion score and better noise removal compared to both the raw, unprocessed input and TinyDenoiser.
\squishend

This paper demonstrates that effective speech AI models can indeed run on low-power wireless hearables. We believe that this  paves the way for integrating on-device AI models for enhanced hearing into billions of wireless earbuds and  hearing aids, unlocking exciting new opportunities.

%% file: system-1.tex
\section{System Design}

We present our hearable hardware, an efficient streaming neural network  and a hardware-software co-design. 


\subsection{NeuralAids Hardware}




The  hardware has a low-power BLE SoC that controls a tiny AI accelerator to process incoming audio in real-time. 
\subsubsection{Stacked printed circuits boards} 
The systems are distributed across the 5  printed circuit boards (PCBs)  shown in Fig.~\ref{fig:hardware}, referred to as PWR, BT, AI, PERIPH and MIC. These circuits are stacked on top of each other and placed in the hearing aid case. The board functionalities are as follows:

\squishlist

\item {\it PWR:} A 2-layer flex board that houses components for power management, battery charging, and monitoring. It also exposes programming pads for the BLE SoC and AI accelerator. The board is powered by a 3.85V battery pack consisting of four CP1254 batteries (75~mAh each, 300~mAh total).  A Texas Instruments BQ25120 power management and charging IC is integrated to charge the battery and generate a regulated 1.8V supply required to power the device components. This chip can also produce a 3.3V voltage domain on demand via I2C to power indicator LEDs on other circuits. A switch is included to manually turn the device on and off. The programming pads are designed for interfacing with a $2\times7$ array of 1.27 mm pogo pins, with one pad also serving as a charging interface for the battery via a 5V source. The board has a long flexible section with a Molex 2167010209 board-to-board connector at the tip that connects  to the BT board. Since the board exposes programming pads, it is placed along the edge of the case.

\item {\it BT:} A 2-layer rigid board containing an ISP2053, which is a BLE SoC based on nRF5340 that integrates additional components such as capacitors, oscillators and an antenna. A resistor is soldered in one of two possible positions to hardcode the device's side (i.e. left  or right device). A board-to-board connector (Molex 513382674) attaches this circuit to the AI board, enabling power and data  exchange between the two boards. 

\item {\it AI:} A 6-layer rigid board houses the low-power AI accelerator (GreenWaves GAP9). An FPF1204UCX load switch is used to enable the BLE SoC to control power to the AI chip, allowing it to connect and disconnect power as needed. This board connects to the PERIPH circuit.

\item {\it PERIPH:} A 4-layer rigid board contains peripheral devices, including an AP Memory APS256XXN 256~Mbit RAM chip and an SSM6515 ultra-low-power I2S DAC and amplifier. The AI accelerator can control power to the RAM chip using a load switch (FPF1204UCX) when the RAM is not needed. Additionally, this board includes a NOR flash (Macronix MX25UW12845G) and an IMU (BMI323), which are reserved for future research. A right-angled 1 mm pitch connector is used to connect a Phonak Audeo Marvel M Receiver-in-canal (RIC), allowing audio playback into the ear canal. The board connects to the MIC circuit using  Molex 5050661022. 

\item {\it MIC:} A 2-layer flexible microphone array with 3 PDM microphones (TDK T5837) and 2  buttons (Omron B3U-1000P).

\squishend

\subsubsection{Hardware subsystems}
The NeuralAid system consists of three main subsystems: Bluetooth, AI, and Audio.

\squishlist
\item {\it Bluetooth subsystem:}  This is built around the BLE SoC, which is the first chip to boot when the device powers on and manages power delivery to the AI chip. It  has direct access to the buttons on the MIC board and receives system-level interrupts such as wake-up and reset signals. There are two  channels between the BLE and  AI chips: 1)  A 115200 baud UART interface for low-speed, spontaneous communication, and 2)  an I2S interface for high-speed, continuous communication. The BLE SoC enables wireless connectivity with external devices. After waking up the AI chip, it transmits advertising packets and accepts connections, allowing remote control  and data exchange for audio streaming or playback. The BLE SoC is dual-core; we  run time-critical BLE-related tasks on one core, while tasks like interfacing or GPIO control  run on another core. 

\item {\it Audio subsystem:} This  manages audio from microphone capture to playback through the RIC.  Although audio interfaces directly with GAP9, all its Serial Audio Interfaces (SAIs) derive their clocks from the BLE chip. GAP9 manages three SAIs for different purposes: 1) reading audio from microphones, 2) writing audio to speakers, and 3) exchanging audio with the BLE SoC (e.g., for streaming). The BLE SoC provides a 3.072~MHz audio clock, enabling high-quality microphone capture. GAP9’s Smart Filter Unit (SFU) converts incoming PDM samples to PCM using an 8th-order cascaded integrator-comb filter with a 64× decimation ratio and 2 samples per stage. The output is shifted by 24 bits to generate 32-bit PCM samples at 48 kHz, which are then downsampled to 16 kHz via the SFU's resampler block for processing.  For playback, the speaker SAI interface transmits 32-bit PCM samples at 48 kHz via I2S to the DAC. Since internal processing occurs at 16 kHz, the SFU’s resampler upsamples audio back to 48 kHz. The BLE I2S interface relays recorded audio and external playback between the BLE and AI chips.

\item {\it AI subsystem:}  The AI accelerator processes incoming audio in real-time using a neural network. Every 96 audio samples (6~ms at 16~kHz), it performs speech enhancement. First, the system converts 32-bit PCM audio to 16-bit float using a dedicated fixed-to-floating point converter. A pre-emphasis filter with a coefficient of 0.97 removes DC components. The audio is then placed in a ring buffer and an FFT converts it into the frequency domain. The neural network processes the frequency-domain audio, after which an inverse FFT and overlap-add reconstruct the time-domain signal. The output is converted back to 32-bit integers for playback through the RIC. Although the microphone array has three microphones, only one is used in this paper. For efficiency, FFT and inverse FFT run on GAP9's compute cluster, while other operations execute on the fabric controller. All GAP9 components, including the fabric controller, compute cluster, and peripherals, are clocked at at 370~MHz. The AI chip also communicates with onboard RAM via an Octal SPI interface.


\squishend

\begin{table}[t]
  \caption{Power consumption of various NeuralAid components. \textmd{Speaker amplification was calibrated by placing a NeuralAid  in a silicone ear model and measuring the sound level  8 mm from the RIC tip. The amplification was adjusted to increase the sound level by 20 dB. The ambient noise level was 33 dBA in both the calibration and testing environments.}}
  \label{tab:power-consumption}
  \vskip -0.15in
  \centering
  \begin{tabular}{ l c }
    \toprule
    \multicolumn{1}{l}{\textbf{Component}} &
    \multicolumn{1}{c}{\textbf{Power (mW)}}\\
    \midrule
    BLE chip &  6.75 \\
    Microphone array &  2.02 \\
    Speaker &  1.49 \\
    \bottomrule
  \end{tabular}
   \vskip -0.05in
 \label{tab:powerconsumption_comp}
\end{table}

\subsubsection{Component power consumption.} Table~\ref{tab:powerconsumption_comp} presents the power consumption of the BLE SoC, microphone array, and speaker while running real-time AI speech enhancement. The combined power consumption of these components is 10.26~mW. The power consumption of the AI accelerator, which varies based on the complexity of the neural network, is analyzed in detail in the following sections.

\begin{figure*}[t!]
\centering
  \includegraphics[width=0.85\textwidth]{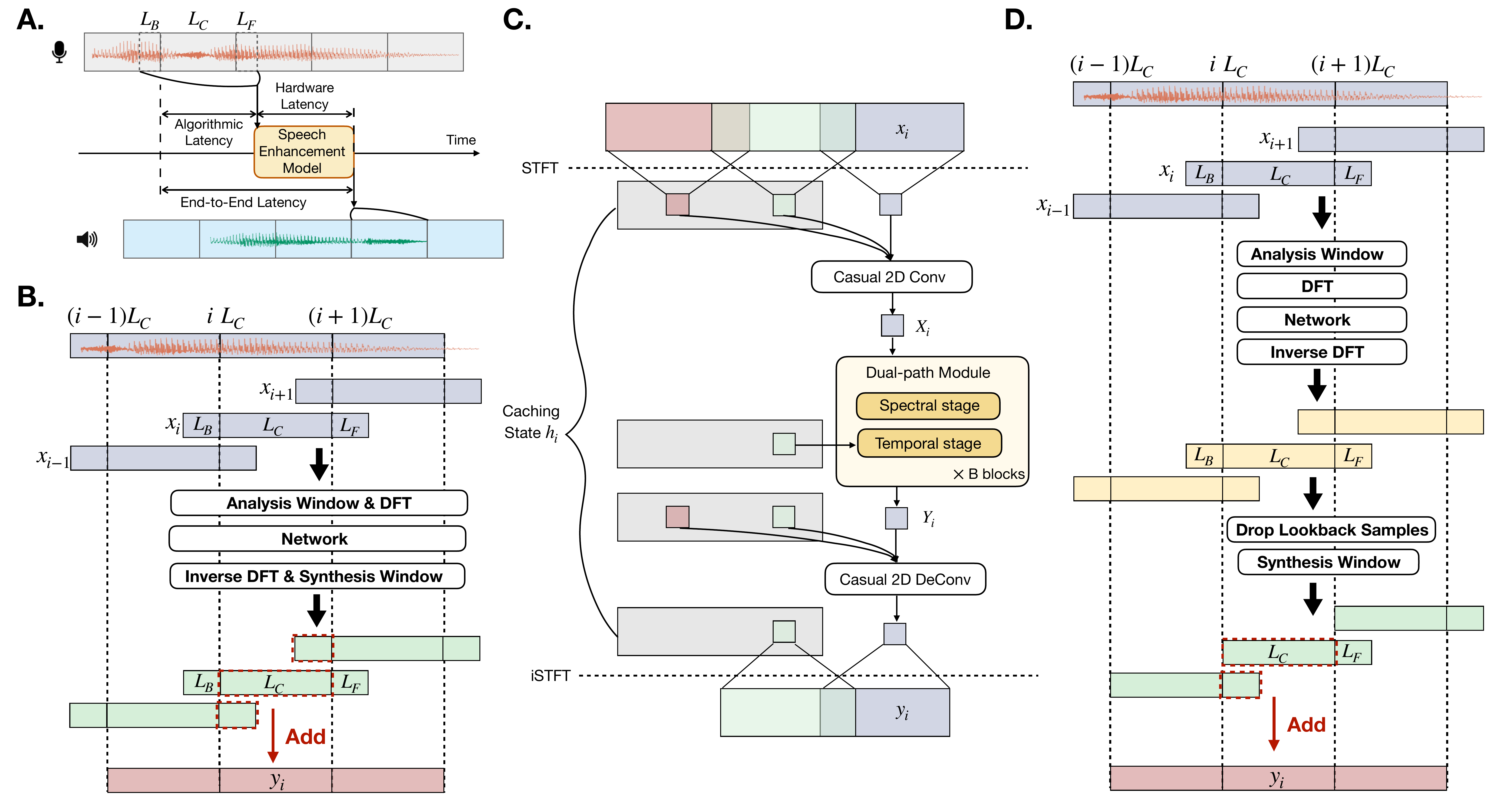}
  \vskip -0.1in
  \caption{{\bf Efficient streaming neural network.} \textmd{ (A) Decomposition of the end-to-end latency in streaming speech enhancement. (B) The normal overlap-add operation introduces additional algorithmic latency with lookback padding. (C) The overall architecture of our streaming speech enhancement network with history caching states for computing reuse. (D) Dual-window approach during overlap-add can reduce the additional algorithmic latency introduced by the lookback padding. }}
  \vskip -0.05in
  \label{fig:network}
\end{figure*}

\subsection{Efficient Streaming Neural Network}\label{sec:nn}

We have three key requirements for achieving a streaming speech enhancement network: (1) achieving a low end-to-end latency of less than 20 ms between the audio and visual scenes, with an algorithmic latency of 10 ms or less, (2) completing the processing of the current chunk before the next chunk arrives, and (3) generalizing effectively to unseen reverberant and noisy environments, as well as diverse wearers, without requiring training data collected from hardware.

As shown in Fig.~\ref{fig:network}A, end-to-end latency is defined as the time for a single audio sample to travel from the microphone input buffer, through the speech enhancement network, and into the speaker output buffer. This latency consists of two components: algorithmic latency, introduced by factors such as chunk size, lookahead, and overlap-add; and 
hardware latency, which accounts for the computation and I/O time required to process each chunk~\cite{wang2022stft}.

\subsubsection{Problem Formulation}
Given an acoustic environment, let $s(t)$ represent the clean target speech received at the microphone, and let $n(t)$ denote all background noise. The mixture audio, $x(t)$, can be expressed as:
\begin{equation}
x(t) = s(t) + n(t)
\end{equation}
Since our  enhancement network $\mathcal{N}$ processes audio chunks in a streaming manner, we divide the continuous waveform into smaller audio chunks: $x(t) = [x_0, x_1, \dots, x_N]$, $s(t) = [s_0, s_1, \dots, s_N]$, and $n(t) = [n_0, n_1, \dots, n_N]$. The process of streaming speech enhancement can be formulated as, 
\begin{equation}
\hat{s}_i, h_i = \mathcal{N}(x_i, h_{i-1})
\end{equation}
Here, $\hat{s}_i$ represents an estimation of the clean speech $s_i$, while $h_{i-1}$ denotes the cache or memory state from previous chunks, which can be reused as the previous intermediate output, and  $h_i$ represents the current cache or memory state.




\subsubsection{Neural Network Architecture}\label{sec:nnchanges}
Existing low-power, on-device speech enhancement networks, like  TinyLSTM~\cite{tinylstms} and its successor TinyDenoiser~\cite{tinydenoiser}, have two  drawbacks: (1) their simple  architecture significantly limits enhancement performance, and (2) they exhibit high algorithmic latency of around 25 ms, restricting real-time applications. While several advanced speech enhancement models have been proposed in recent years~\cite{tfgridnet, chen2020dual, tfloco}, they fail to meet our real-time, on-device, and low-power constraints. Our goal is to significantly improve speech enhancement performance while maintaining low-latency and real-time constraints.

The overall architecture of our network is shown in Fig.\ref{fig:network}C. It is a dual-path Time-Frequency (TF) domain model based on TF-GridNet\cite{tfgridnet}, which  achieves superior performance for  speech enhancement  tasks. First, the small audio chunk  $x_i$  is transformed into its time-frequency representation $X_i$ using a Short-Time Fourier Transform (STFT) followed by a causal 2D convolutional layer. The transformed representation $X_i$ is then processed by multiple dual-path modules—the most computationally intensive components of the network. Each dual-path module has two stages: a spectral stage that processes frequency information and a temporal stage that handles time dependencies. Finally, the enhanced time-frequency representation  $Y_i$ is passed through a 2D causal deconvolution layer and reconstructed back into a time-domain signal $y_i$  using the inverse STFT and overlap-add operations.


To enable real-time operation on low-power devices while maintaining low latency, we  highlight three  modifications.

\vskip 0.05in\noindent\textbf{Low-complexity dual-path architecture.} In the original TF-GridNet, the spectral stage consists of a bi-directional LSTM that processes the frequency domain of $X_i$, while the temporal stage consists of a causal uni-directional LSTM that processes the time domain of $X_i$. We observed that the primary contributor to inference time is the bi-directional LSTM in the spectral stage, as it must process the long sequence of frequency bins sequentially. Using recurrent models to process frequency sequences sequentially slows computation. 

To reduce its complexity, we first apply a strided convolution layer to compress the frequency dimension before using the frequency-domain LSTM. This reduces the sequence length processed by the LSTM, thereby decreasing its execution time. Additionally, compared to LSTMs, GRUs offer lower computational cost while maintaining similar performance~\cite{chung2014empirical}. To further optimize the runtime of the spectral module, we replace the LSTM module with a GRU module.

\vskip 0.05in\noindent\textbf{Dual-window approach for low-algorithmic latency.}
As shown in Fig.~\ref{fig:network}A, to achieve  low algorithmic latency, we process the input audio $x_i$ in small chunks of size $L_C$ (ms), with an additional lookahead of $L_F$ (ms) and a lookback of $L_B$ (ms); we set $L_C = 6 ms$, $L_B=6 ms$ and $L_F=4 ms$. Lookahead and lookback padding improve frequency resolution, which is important for enhancement performance \cite{wang2022stft}. However, increasing the lookahead and lookback padding also introduce higher algorithmic latency.

With lookahead padding $L_F$, the network must wait for an additional $L_F$  ms of audio samples before applying the STFT. While lookback padding does not introduce additional waiting time at input, it causes algorithmic latency during the iSTFT process. Specifically, the overlap-add processing step of the current chunk $y_i$ requires the lookback padding from the next chunk $y_{i+1}$, as shown in Fig.~\ref{fig:network}B. Thus, the overlap-add process must wait for the next chunk (an additional $L_B + L_C$  ms) to reconstruct the current output chunk.

 To eliminate the algorithmic latency introduced by lookback padding, we adopt a dual-window-size approach~\cite{wang2022stft, mauler2007low}, where the synthesis window is shorter than the analysis window, as shown in Fig.~\ref{fig:network}D. By discarding the lookback samples after the inverse DFT and before applying the synthesis window, the overlap-add process for the current chunk no longer depends on the next chunk’s output. We use a rectangular window for analysis and use a specific function for the synthesis window $s$, described below, valid for $L_C > L_F$, to achieve perfect signal reconstruction~\cite{griffin1984signal}:  
  \[s[i] = \begin{cases} 
 1 & \text{if } i \in [L_F, L_C]\\
 \frac{1}{\floor{\frac{L_C + L_B}{L_C}} + 1} & \text{otherwise }
 \end{cases}\]
With this approach, our lookback padding can increase frequency resolution without introducing additional algorithmic latency. As a result, we achieve a low algorithmic latency of  $L_c + L_F = 10~ms$.


\vskip 0.05in\noindent\textbf{Cache state management.} When processing a stream of continuous chunks, many values can be reused to avoid redundant computations. To achieve this, we maintain and update a cache state $h_i$ at every inference step. As shown in Fig.~\ref{fig:network}C, our cache state $h_i$ consists of four components: 1)   Since the kernel of the first 2D causal convolution layer requires prior chunks, we can avoid recomputing the STFT frames of these prior chunks by caching and reusing them, as illustrated in the figure. 2)   We store the output of the sequence of dual-path modules from previous chunks, allowing us to compute the causal 2D deconvolution more efficiently. 3)  Since computing the ISTFT for the current chunk  depends on previous chunks, we maintain a buffer for the intermediate outputs of the 2D deconvolution layer. 4) We store the hidden and cell states of the temporal unidirectional LSTM for each dual-path module. This allows us to fully leverage the long-term receptive field of the recurrent network.

\subsubsection{Training for Real-world Generalization}\label{sec:training}
Our speech enhancement system must generalize to complex real-world acoustics, where variations arise from multipath propagation, head-related transfer functions (HRTFs), diverse noise profiles, and motion. To ensure real-world generalization, we adopt a two-step training strategy.


\noindent\textbf{Training.} In the first stage, we train our model on the LibriSpeech~\cite{librispeech}  dataset (360 hours of clean speech).  To enhance generalization across reverberation and human wearers, we convolve each speech sample and background noise with a binaural room-impulse-response (BRIR) that captures the acoustic transformations caused by a room  and the human head.
We use four real-world BRIR datasets—CIPIC\cite{CIPIC}, RRBRIR~\cite{rrbrir}, ASH-Listening-Set~\cite{ASH}, and CATTRIR~\cite{CATT_RIR}— and split them into training, validation, and test sets without overlap. We also use WHAM! (58.03 hours of diverse background noise)~\cite{wichern2019wham}. For each training sample, we randomly select  a 5-second clean speech clip $a(t)$ from LibriSpeech, a BRIR $h_{\theta, \phi}(t)$ from  BRIR datasets recorded at azimuthal angle $\theta$ and polar angle $\phi$, and a 5-second background noise clip $n(t)$ from WHAM. Then we mix them as, $x(t) = h_{\theta, \phi}(t) \ast a(t) + n(t)$.  We do not convolute $n(t)$ with $h_{\theta, \phi}(t)$, as WHAM!  noise  is recorded in a binaural format.

During  training, our  target is  $s(t) = h_{\theta, \phi}(t) \ast a(t)$ to  preserve the binaural characteristic of clean speech. We optimize the model using the SNR loss:  $ L_{SNR}\Big(\hat{s}, s\Big) =   \frac{\| s \|^2}{\| s - \hat{s} \|^2}$.   Training runs for 200 epochs, each with 20k samples, using AdamW with gradient clipping (0.1). The learning rate follows three sequential schedulers: 1)  linearly increases from 1e-4 to 1e-3 over 10 epochs, 2) maintains 1e-3 for 140 epochs, and 3) halves every 15 epochs for the final 50 epochs.

In practice, each device runs its own network, processing audio independently. Thus, we train two separate networks — one for the left channel and one for the right.

\noindent\textbf{Fine-tuning with motion and data augmentation.} In the previous  stage,   the sound source and wearer's head were assumed to be  static. However, real-world scenarios involve motion. 
 To address this, we fine-tune the model with time-varying $\theta$ and $\phi$, following  motion simulation  from ~\cite{lookoncetohear}.

 Specifically, the source’s position updates every 25 ms, with a 2.5\% probability of triggering a random motion event. When triggered, the angular velocity in the azimuthal and polar directions is sampled from [$\pi/6, \pi/2$] rad/s, and the source moves at this velocity for a random duration between 0.1 and 1 s, during which no other motion event is triggered. The Steam Audio SDK~\cite{steamaudio-sdk} is used for motion trajectory simulation, creating a mix of stationary and moving segments within the same audio clip. Since BRIR datasets contain recordings at discrete positions, we approximate intermediate positions using nearest-neighbor BRIR selection based on the current $[\theta(t), \phi(t)]$.

We also augment, with a 30~\% probability, the mixture signal with white, pink, and brown noise, mimicking real-world noise sources like microphone thermal noise and HVAC systems. White noise is generated with a standard deviation sampled from [0,0.002), while pink and brown noise are created using the Python \verb|colorednoise| library and scaled by a random factor from [0,0.05]. Additionally, also with a 30~\% probability, we randomly augment the speed of the target speech to 80--120~\% of its original value. We fine-tune the model for 100 additional epochs, incorporating motion simulation and noise augmentation. Instead of SNR loss, we use a multi-resolution spectrogram loss~\cite{multi-resolution, auraloss}. 
 Optimization uses AdamW with a gradient clipping of 0.1, and a \verb|ReduceLROnPlateau|  scheduler (patience = 5, factor = 0.5) starting at 1e-3 learning rate.

\subsection{Hardware-Software Co-Design}

\subsubsection{Hardware Constraints.} GAP9's limited memory and compute resources impose strict constraints on neural networks that it can run efficiently. It has only 1.5~MB of L2 memory and 128~KB of L1 memory. Models exceeding this require external L3 memory, which is slow and power-intensive. 

Efficient DSP and neural network inference rely on GAP9's compute cluster — 9 RISC-V cores and the NE16 accelerator. The NE16 is optimized for streamed multiply-accumulate operations but only supports up to 8-bit quantized weights and 8-/16-bit activations. The RISC-V cores handle both integer and floating-point operations but are less efficient for neural networks. Thus, the network must be carefully designed by identifying the parts of the network that can be efficiently run on the NE16 with quantization without degrading the output speech quality.

\subsubsection{Quantizing Neural Networks. }


Quantization is the process of converting  tensors in the computational graph from a full-precision representation (e.g., \textsc{fp32}) to a lower-precision fixed-point representation (e.g., \textsc{int8}). This helps reduce memory usage, runtime and power consumption. 

We denote the parameters of the network $\mathcal{N}(\cdot)$ as $\theta$, which are originally stored in floating-point precision. Quantization can be classified into two types:
\squishlist
\item  {\it Weight quantization:} Applied to  network parameters $\theta$.
\item {\it Activation quantization:} Applied to  intermediate activation maps generated during inference.
\squishend

We use uniform \textsc{int8} quantization, with asymmetric thresholds for activations and symmetric thresholds for weights.


A uniform quantizer is defined as follows: let $r \in \mathbb{R}^n$ be a vector to be quantized, $S \in \mathbb{R}^+$ be the quantizer scaling factor, $ Z \in \mathbb{R}$ be the zero-point and let $b$ be the bit width. The quantization process  can be formulated as:
\begin{equation}
Q(r) = \lfloor r/S \rceil + Z
\end{equation}
where $Q(r)$ is the fixed point representation of $r$ in quantization space. $\lfloor \cdot \rceil$ denotes the rounding of the input to the nearest integer value.

Dequantization  maps   $Q(r)$ back to $r$, and is given by:
\begin{equation}
\hat{r} = S(Q(r) - Z)
\end{equation}
Since the recovered values $\hat{r}$ may not exactly match $r$ due to rounding, a {\bf quantization error} is introduced.


A key aspect of uniform quantization is choosing the  scaling factor $S$. It defines how a given range $[\alpha, \beta]$ of real values $r$  is divided into discrete partitions~\cite{gholami2022survey}:
$S = \frac{\beta - \alpha}{2^b - 1}$. 
 In our implementation, we use a Min-Max Moving Average Observer to determine $[\alpha, \beta]$.

In asymmetric quantization, the  range  $[\alpha, \beta]$ is not necessarily symmetric with respect to the origin, i.e., $-\alpha \neq \beta$. In symmetric quantization, in contrast, a symmetric clipping range  ($-\alpha=\beta$) is used and its scaling factor $S$ is computed as, $S = \frac{\text{max}(-\alpha, \beta)}{2^{b-1} - 1}$~\cite{gholami2022survey},  with the zero-point set to $Z = 0$.

Furthermore, in  neural networks,  quantization is usually applied to a high-dimension tensor (dimension $\geq$ 2). There are two  methods for quantizing higher dimension tensors: 
\squishlist
\item  {\it Per-tensor quantization:} The entire tensor is quantized using the same $S$ and $Z$.
\item {\it Per-channel quantization:} Each channel is quantized independently with different $S$ and $Z$. 
\squishend
Given the constraints of the GAP9 hardware, we deploy per-tensor asymmetric quantization for activations and per-channel symmetric quantization for weights.

\subsubsection{Mixed-Precision Configuration}\label{sec:mixedprecision}
One  method for quantizing a trained floating-point neural network is Post-Training Quantization (PTQ). This involves: 1) calibrating the quantization range using a representative dataset, and 2) quantizing all tensors in the network to a lower-precision format.

We start PTQ by applying full \textsc{int8} quantization, but as shown in Table~\ref{tab:main-qat}, this results in severe performance degradation due to quantization errors. Instead, inspired by~\cite{tinydenoiser11}, we adopt mixed-precision quantization to mitigate performance loss. The key idea is that different network submodules have varying sensitivity to quantization errors. By selectively quantizing certain modules to \textsc{bfloat16} instead of \textsc{int8}, we can preserve critical information and recover performance from quantization errors. According to ~\cite{fqss, tinydenoiser11},  input and output quantization is sensitive to the quantization error and leads to performance degradation. The rationale is that the first convolution layer processes raw input data, so maintaining high precision helps retain essential features.
The last deconvolution layer reconstructs the final output, where high precision is crucial for generating high-quality audio.  
Hence, we quantized first input convolution and last deconvolution layers  to \textsc{bfloat16} instead of \textsc{int8}. 

\subsubsection{Quantization-Aware Training for Better Performance} 

While mixed-precision PTQ helps mitigate some performance loss, there remains a significant performance drop compared to the floating-point network. This degradation is even more pronounced (Table.~\ref{tab:main-qat}) in  model architectures with a large number of sub-components like ours,  which restricts the deployment of advanced models on our accelerator.

To further reduce quantization errors, we apply Quantization Aware Training (QAT). The key idea behind QAT is to { simulate quantization errors during training}, allowing the model to adjust and converge to a more optimal solution under quantization constraints.

During QAT, floating-point weights and activations are rounded to their quantized equivalents. Since quantization is a non-differentiable operation, we use the Straight-Through Estimator  to approximate gradients, where: $\partial \lfloor x \rceil / x = 1$.

\vskip 0.05in\noindent{\bf QAT implementation.} We describe the details below.

    {\it Initialization:} We begin with a pretrained floating-point model with parameters $\theta$ and apply the mixed-precision quantization described in~\xref{sec:mixedprecision}.
    
    {\it Learned Step-Size Quantization (LSQ):} We enhance vanilla QAT with LSQ~\cite{lsq}, which allows the network to learn the quantizer scale factor $S$ using gradients of the task loss. This helps reduce activation quantization errors.
    
    {\it Fine-tuning:} The mixed-precision model is fine-tuned for 30 epochs using the FQSE framework~\cite{cohen23_interspeech}.
    
    {\it Training Setup:}  Each epoch processes only 4,000 mixtures due to QAT’s high computational cost. We use an initial learning rate of 1e-3, with a \verb|ReduceLROnPlateau|  scheduler. 


%% file: results-1.tex
\section{Experiments and Results}

\subsection{Benchmark results}\label{sec:benchmarks} 
\subsubsection{Evaluation Metrics.}
Our evaluation consists of two main components: speech enhancement evaluation and system evaluation. The former assesses the network’s noise suppression capability and the quality of the enhanced speech.

\squishlist
\item {\it SISDRi:} Scale-Invariant Signal-to-Distortion Ratio (SISDR) is a commonly used metric in speech enhancement tasks to assess the quality of enhanced speech relative to the reference clean speech~\cite{le2019sdr}. SISDR Improvement (SISDRi) can be computed between the input and output speech to assess how much the network improves the quality of the speech compared to the clean reference speech.

\item {\it PESQ:} Perceptual Evaluation of Speech Quality (PESQ) is an objective metric that assesses how closely an enhanced signal matches a reference clean speech signal~\cite{pesq}, mimicking human perception.
\item {\it DNSMOS:} Deep Noise Suppression MOS (DNSMOS) is a neural network based objective metric designed to predict Mean Opinion Score (MOS) for speech quality in speech enhancement. We use the  ITU P.835 personalized DNSMOS OVRL implementation as our score~\cite{dnsmos}.
\squishend

For system evaluation, we measure memory consumption, runtime, and power consumption on the AI accelarator.
\squishlist
\item {\it Memory:} GreenWaves' conversion process reports where neural network parameters are stored. We sum L1 and L2 memory usage to determine total memory consumption.
\item {\it Runtime:} Models run on GAP9 by sending tasks to its compute cluster. Runtime is measured by sending a model execution task, counting elapsed clock cycles until the task completes, and dividing the elapsed cycles by clock frequency. The reported runtime is averaged over 100 iterations.
\item {\it Power:} To measure GAP9's power consumption during continuous speech enhancement, we power it with a 1.8V source, run inference on each incoming audio chunk every 6 ms, and measure the current over several minutes. Power is calculated as the product of current and voltage.
\squishend

\subsubsection{Model comparison.}
We compare multiple FP32 models that when fully quantized meet our real-time requirements: 
\squishlist
\item {\it TinyDenoiser:} Prior state-of-the-art TinyDenoiser model \cite{tinydenoiser11} performs on-device speech denoising. For fair comparison, we modify its algorithmic latency to 10ms with the same chunk size and padding as our model. 
\item {\it TFGridNet-6F:} We use the causal implementation from~\cite{lookoncetohear} without self-attention and LayerNormalization modules. Its hyperparameters are $L_B = 6ms$, $L_C = 6ms$, $L_F = 4ms$, $B=6$, $D=32$ and $H=32$. Since TF-GridNet is not real-time even at 4× compression, we apply 6× frequency compression to the spectral module to meet real-time requirements.
\item {\it Our Model:} We set model hyperparameters to $L_B = 6ms$, $L_C = 6ms$, $L_F = 4ms$, $B=6$, $D=32$ and $H=32$. We apply 4× frequency compression on the spectral module.
\squishend
As shown in Table.~\ref{tab:fp32results}, our FP32 model achieves the best SISDRi, PESQ and DNSMOS scores at floating model resolution.

\begin{table}[t!]
  \caption{Floating-point  (FP32) network results.}
  \label{tab:fp32results}
  \vskip -0.1in
  \centering
\setlength{\tabcolsep}{2.4pt}
 
  \begin{tabular}{ l c c c }
    \toprule
    \multicolumn{1}{l}{{Model}} &
    \multicolumn{1}{c}{{SISDRi (dB)}} & 
    \multicolumn{1}{c}{{PESQ}} &
    \multicolumn{1}{c}{{DNSMOS}} \\
    \midrule
     TinyDenoiser & 6.24 $\pm$ 3.43 &  1.37 $\pm$ 0.27  & 2.06 $\pm$ 0.64  \\ 
    TFGridNet-6F & 8.43 $\pm$ 3.46 &  1.74 $\pm$ 0.49  & 2.48 $\pm$ 0.68 \\
    Our model  & 8.65 $\pm$ 3.41  & 1.76 $\pm$ 0.49  & 2.50 $\pm$  0.66  \\
    \bottomrule
  \end{tabular}
   \vskip -0.2in
\end{table}

\begin{table*}[t!]
  \caption{Power consumption and quantization results. \textmd{We measure the SISDRi, model size, runtime and power consumption for different Post-Training Quantization (PTQ) and Quantization-Aware Training (QAT)  strategies and different models. The \textsc{bfloat16} network configurations cannot run in real-time on the target AI accelerator.}}
  \label{tab:main-qat}
  \vskip -0.1in
  \centering

  \begin{tabular}{ l c c c c c }
    \toprule
    \multicolumn{1}{l}{\textbf{Model}} &
    \multicolumn{1}{l}{\textbf{Quantization config}} & 
    \multicolumn{1}{c}{\textbf{SISDRi (dB)}} &
    \multicolumn{1}{c}{\textbf{Memory (kB)}} &
    \multicolumn{1}{c}{\textbf{Runtime (ms)}}&
    \multicolumn{1}{c}{\textbf{Power (mW)}}\\
    \midrule
    TinyDenoiser & \textsc{BFLOAT16}  & 6.30 $\pm$ 3.52  & -- & -- & --\\
    
    TinyDenoiser & \textsc{int8} PTQ & 3.54 $\pm$ 3.34 & 1135.2 & 0.53 & 23.08 \\ 
    TinyDenoiser  & Mix PTQ & 3.90 $\pm$  3.33 & 1195.9  & 0.58 & 24.12\\ 
    TinyDenoiser &Mix QAT &  5.97 $\pm$ 3.35 & 1195.9 & 0.58 & 24.12 \\ 
    \midrule
    Our model & \textsc{BFLOAT16}  & 8.76 $\pm$ 3.41 & -- & -- & -- \\ 
    Our model & \textsc{int8} PTQ & -1.70$\pm$  7.5  &   280.4 & 5.19 & 58.57 \\ 
    Our model & Mix PTQ  & 0.90 $\pm$ 5.40 & 298.8 & 5.54 & 71.64 \\ 
    Our model & Mix QAT & 8.19 $\pm$ 3.38 & 298.8  & 5.54 & 71.64 \\ 
    \bottomrule
  \end{tabular}
\end{table*}

\subsubsection{Quantization evaluation.}\label{sec:quantresults}
Next, we evaluate how different quantization configurations  affect  performance, runtime, power, and memory consumption. We  compare our model with TinyDenoiser~\cite{tinydenoiser11} under three  configurations:
\squishlist
\item {\it \textsc{bfloat16}:} We quantize both TinyDenoiser and our model into \textsc{bfloat16}  resolution.
\item {\it \textsc{int8} PTQ:} We quantize both TinyDenoiser and our model into \textsc{int8} resolution, using Post Training Quantization.
\item {\it Mix PTQ:} For TinyDenoiser, we follow the mix-precision configuration as the original paper~\cite{tinydenoiser11}, where we quantize  the linear layers into \textsc{bfloat16}  and quantize the LSTM layers into \textsc{int8}. For our model, we quantize the input 2D Conv and output 2D DeConv into \textsc{bfloat16} and quantize other parts into \textsc{int8}. Then we apply Post Training Quantization on these mixed precision models.
\item {\it Mix QAT:} We follow the same mixed precision quantization configuration for TinyDenoiser and our model, then we apply Quantization-Aware Training. 
\squishend

The results of different quantization configurations are shown in Table~\ref{tab:main-qat}. While full INT8 quantization achieves the lowest runtime, memory usage, and power consumption for both TinyDenoiser and our model, it significantly degrades speech enhancement performance. TinyDenoiser experiences a 2.76~dB drop, while our model suffers a 10.46~dB decline. The larger performance drop in our model is due to more severe quantization noise accumulation in its deeper and more complex architecture.

Applying mixed-precision quantization helps recover 0.36~dB of SISDRi for TinyDenoiser and 2.6~dB for our model. However, the performance loss compared to the floating-point model remains non-negligible. After QAT, TinyDenoiser achieves an SISDRi of 5.97~dB, while our model reaches 8.19~dB, demonstrating the effectiveness of QAT, especially for deeper and more complex networks. Our mixed-precision model with QAT achieves the best performance within hardware constraints. Considering runtime, memory, and power consumption, our model balances performance and memory efficiency while maintaining real-time requirements ($< 6~ms$) and power constraints ($71.6~mW$), trading some runtime and power efficiency for improved speech enhancement.

\begin{figure}[t!]
\centering
\vskip -0.15in
\includegraphics[width=0.7\linewidth]{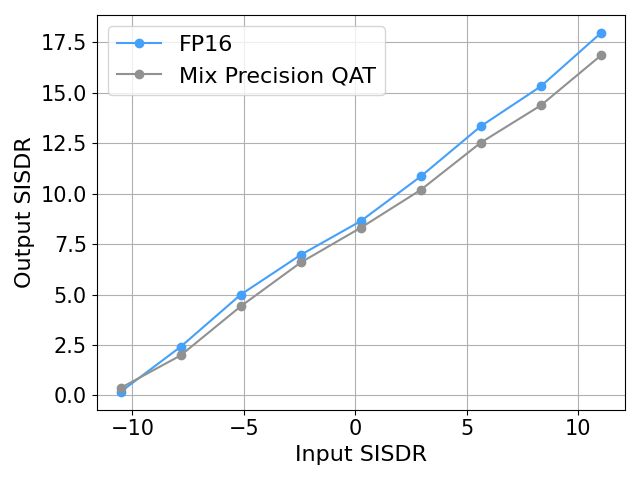}
\vskip -0.15in
\caption{QAT reduces the  gap between our  floating-point and quantized models across  input noise levels.}
\label{fig:in_out}
\vskip -0.19in
\end{figure}

In Fig.~\ref{fig:in_out}, we plot the output SISDR as a function of input SISDR for both the floating-point model and the mixed-precision QAT model. The results show that QAT reduces the performance gap between our  floating-point and quantized models across all input noise levels. An interesting observation is that the performance gap increases with higher input SISDR. As input SISDR increases, the output speech quality becomes clearer, making quantization noise more dominant and its effects more pronounced compared to scenarios with lower input SISDR~\cite{cohen23_interspeech}.

\subsubsection{End-to-end hardware runtime evaluation.} 
\begin{figure}[t!]
    \centering
    \begin{subfigure}[t]{0.24\textwidth}
        \centering
        \includegraphics[width=0.95\textwidth]{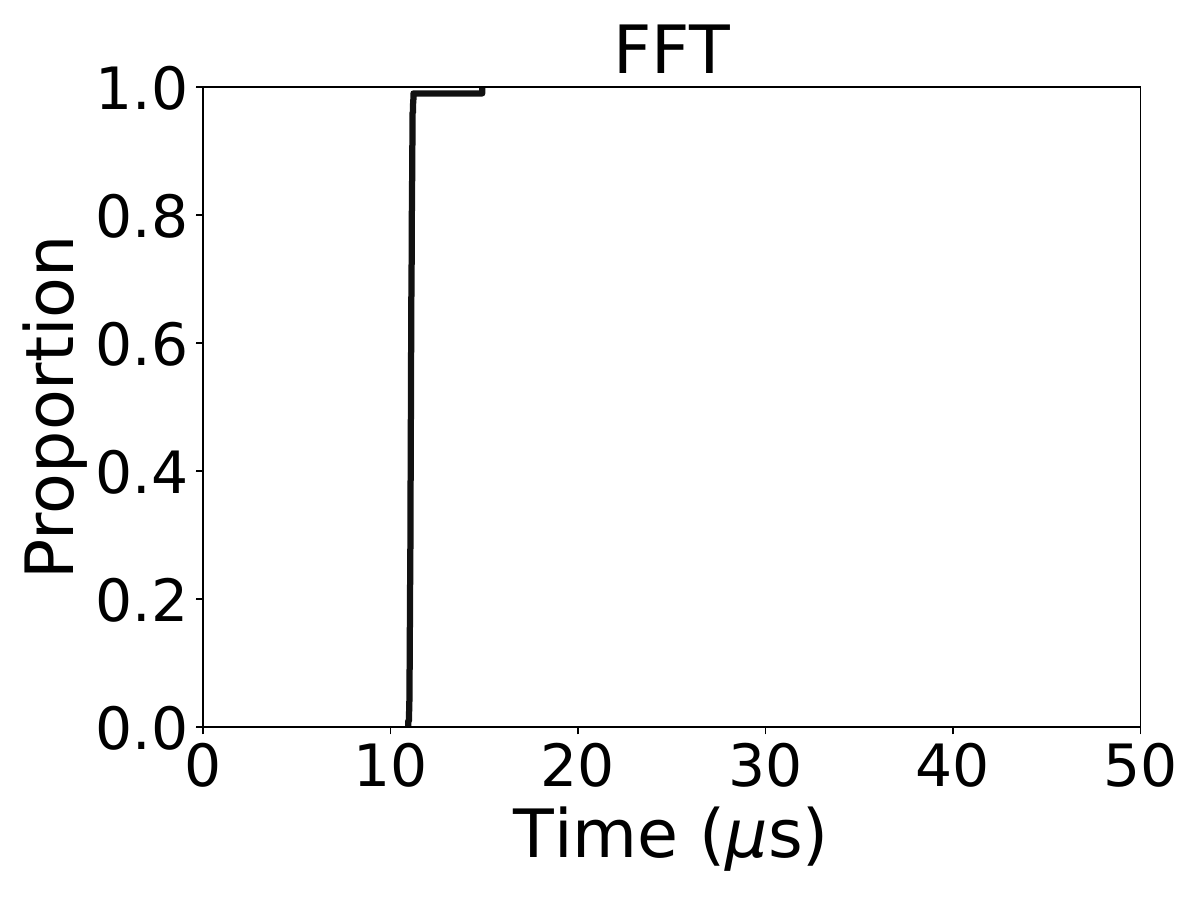}
        \vskip -0.1in
        \caption{}
        \label{fig:fft_runtime}
    \end{subfigure}
    ~
    \begin{subfigure}[t]{0.24\textwidth}
        \centering
        \includegraphics[width=0.95\textwidth]{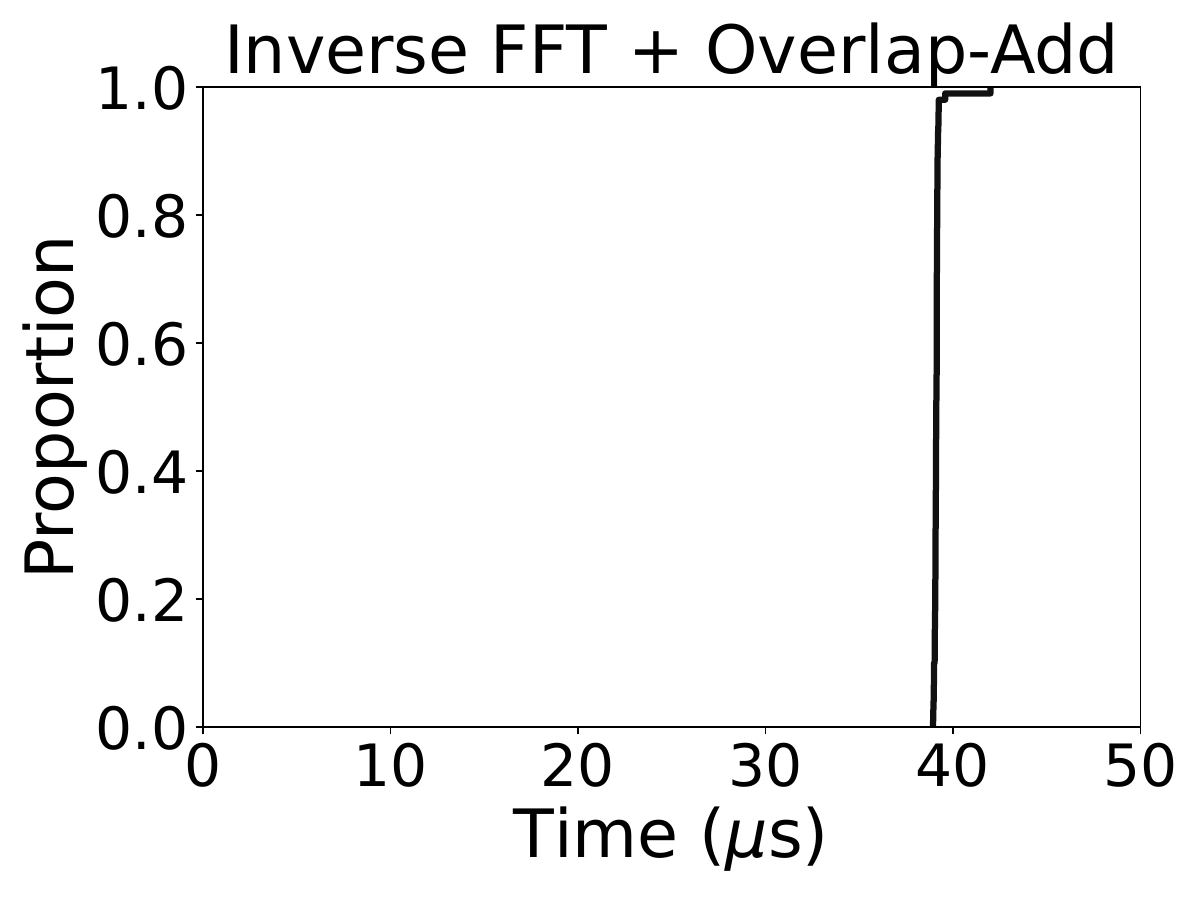}
                \vskip -0.1in
        \caption{}
        \label{fig:ifft_runtime}
    \end{subfigure}
    
    \begin{subfigure}[t]{0.24\textwidth}
        \centering
        \includegraphics[width=0.98\textwidth]{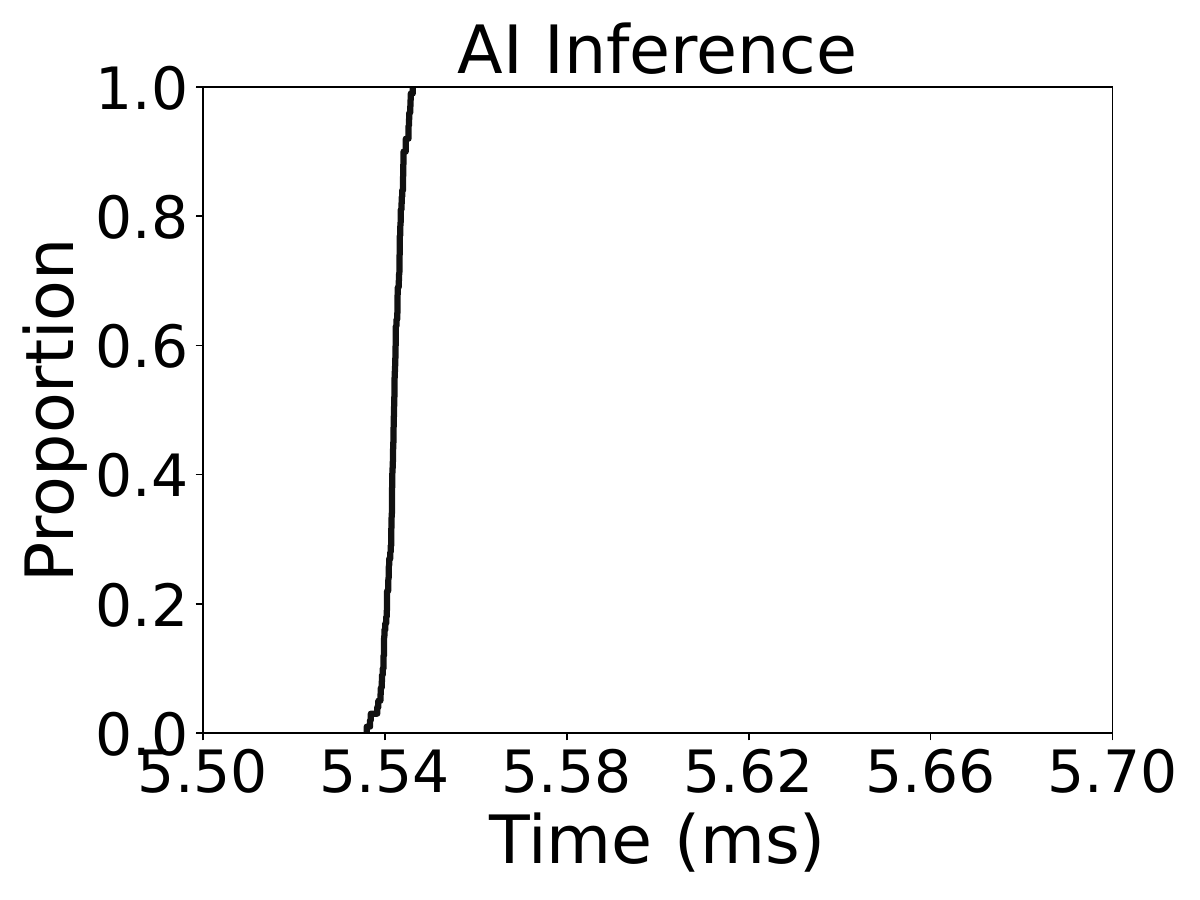}
                \vskip -0.1in
        \caption{}
        \label{fig:ai_runtime}
    \end{subfigure}
    ~
    \begin{subfigure}[t]{0.24\textwidth}
        \centering
        \includegraphics[width=0.98\textwidth]{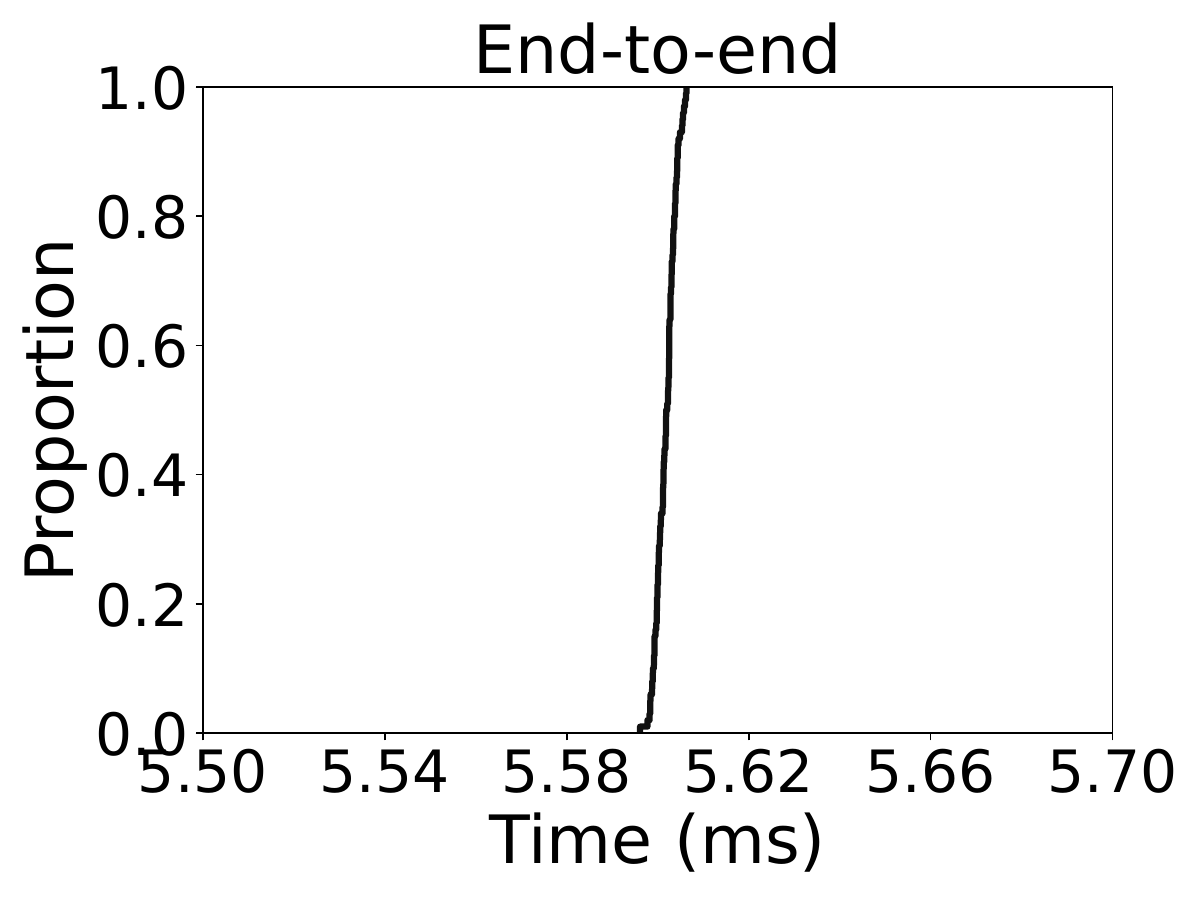}
                \vskip -0.1in
        \caption{}
        \label{fig:e2e_runtime}
    \end{subfigure}
    \vskip -0.15in
    \caption{End-to-end hardware run-time evaluation.}
    \vskip -0.15in
    \label{fig:runtime}
\end{figure}

Fig.~\ref{fig:runtime} shows CDF plots of the hardware runtime for key subcomponents of our audio processing pipeline. In addition to AI inference with our speech enhancement model, the pipeline performs FFT and inverse FFT for time-frequency conversion. Runtime is measured by recording clock cycles before and after the execution of each component and dividing the difference by the clock frequency. The plots are generated by consolidating runtime measurements over 100 iterations. Since we use rectangular synthesis windows (i.e., containing all ones), there is no need to apply a window before the FFT. However, for the inverse FFT, we apply an analysis window (not necessarily all ones) and so include the overlap-add operation in the runtime measurements. Figs.~\ref{fig:fft_runtime}-\ref{fig:ai_runtime} show consistent runtimes, with AI inference being the most time-consuming step. Fig.~\ref{fig:e2e_runtime} shows that the full pipeline, including additional tasks such as data copying and type conversion, executes in under 6 ms, ensuring real-time performance.


\subsubsection{Wireless throughput.}

\begin{figure}[t!]
\centering
\includegraphics[width=0.75\linewidth]{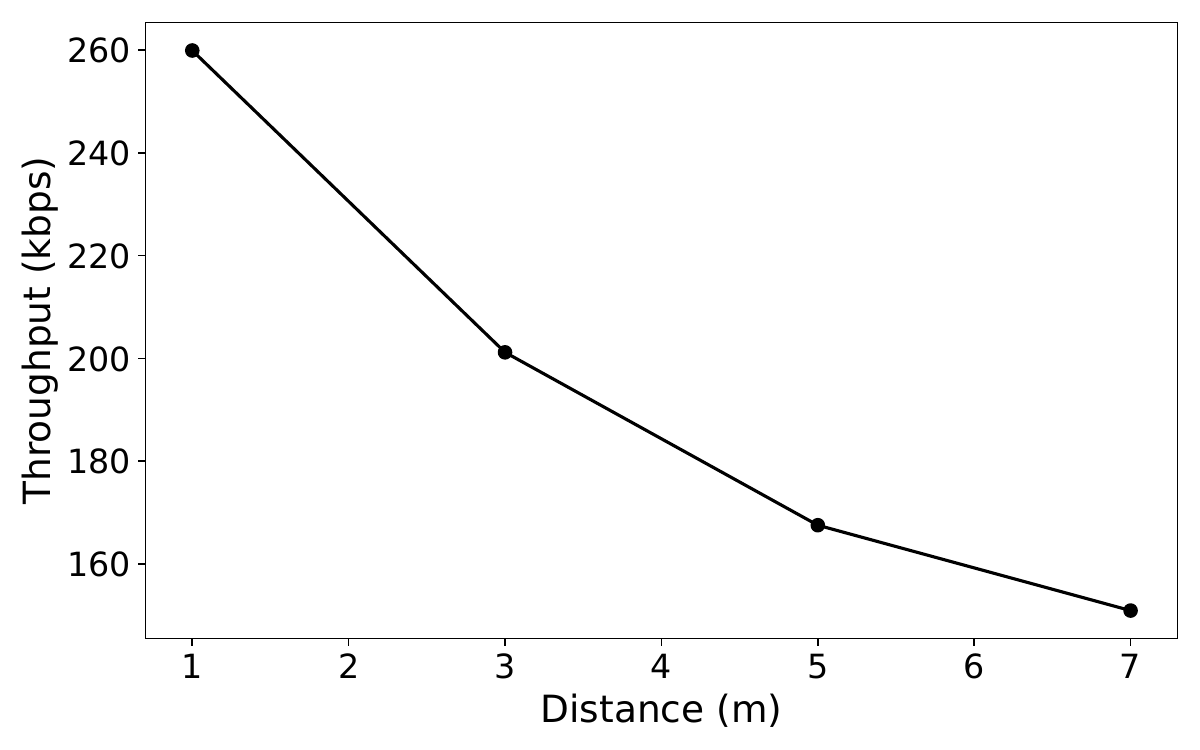}
\vskip -0.15in
\caption{Wireless throughput from the NeuralAid device to a nearby receiver as a function of distance.}
\label{fig:throughput}
\vskip -0.15in
\end{figure}

Fig.~\ref{fig:throughput} plots the wireless throughput of our NeuralAid device at different distances from the receiver. To measure throughput, we stream 10,000 packets, each 196 bytes in size, over BLE from the hearing aid to a laptop in a large conference room. The total transmission time is recorded and used to compute the throughput. As expected, throughput decreases as the distance increases due to higher packet loss and retransmissions. Beyond 3 meters, throughput drops below 200 kbps, primarily due to the limited range of the device’s low-profile integrated antenna.

\subsection{In-the-Wild Evaluation}

\begin{figure*}[t!]
\centering
\includegraphics[width=\linewidth]{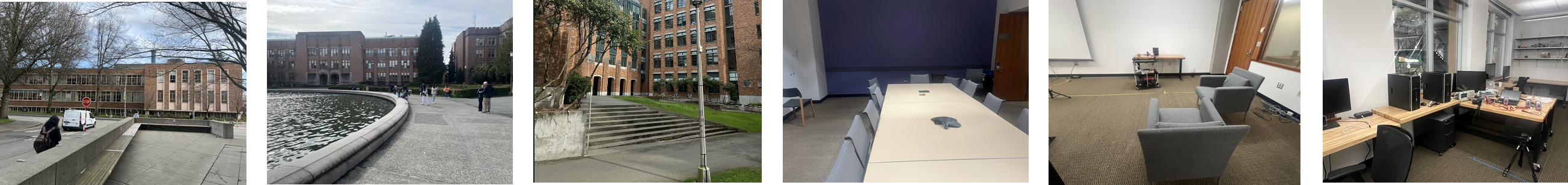}
\vskip -0.11in
\caption{Different outdoor and indoor in-the-wild scenarios. \textmd{ Indoor pictures taken without humans just for this figure.  }}
\label{fig:scenarios}
\vskip -0.05in
\end{figure*}

We evaluated our system in previously unseen indoor and outdoor environments, using participants who were not included in the training data. It is important to emphasize that our training data consists solely of publicly available datasets, as detailed in~\xref{sec:training}, and does not include any data collected with our own hardware. The various acoustic environments evaluated are shown in Fig.~\ref{fig:scenarios}, covering indoor settings such as noisy office spaces, as well as outdoor locations like busy streets with traffic noise and natural environments like parks with multiple noise sources. In indoor spaces, the noise sources included chatter from a nearby event with many people, as well as sounds of coughing, table knocking, and other typical office noises. In outdoor settings, in addition to traffic noise, we also encountered airplane sounds.  In all these settings, we had a speaker read a different text in the presence of uncontrolled environmental sounds. The wearer and the speaker were free to move and/or rotate their head    and adopted different postures like sitting and standing.

{\bf Evaluation procedure.} Since the  speaker is speaking in the presence of  unknown noise, it is difficult to obtain the ground truth audio signal for our  speakers in the real world. So, we cannot rely on objective metrics to evaluate the system performance. Instead, we use subjective metrics to allow human participants to rate the audio quality. 

For this user study, we recruited 28 participants (19 male and 9 female) ranging in age from 18 to over 70 years, with an average age of 43. The only inclusion criteria was that the participants were adults and could follow English instructions. Each participant evaluated the system in three modes, selected in a random order, across 15 scenarios.

\begin{enumerate}
    \item \textbf{No AI}: In this mode, there is no noise suppression and thus the participants hear the unfiltered mixture of the speech signal and background noise.
    \item \textbf{TinyDenoiser}: Prior state-of-the-art TinyDenoiser model. 
    \item \textbf{Our method}: In this mode, we use our NeuralAid model fine-tuned with motion and colored noise as described in~\xref{sec:training}.
\end{enumerate}

For each of these modes, we ask the participants to rate the speech quality by asking them the following questions~\cite{lookoncetohear}:

\begin{enumerate}
    \item \textbf{Noise suppression}: \textit{How INTRUSIVE/NOTICEABLE were the INTERFERING SPEAKERS and BACKGROUND NOISES? 1 - Very intrusive, 2 - Somewhat intrusive, 3 - Noticeable, but not intrusive, 4 - Slightly noticeable, 5 - Not noticeable}
    \item \textbf{Overall MOS}: \textit{If the goal is to focus on this target speaker, how was your OVERALL experience? 1 - Bad, 2 - Poor, 3 - Fair, 4 - Good, 5 - Excellent}
\end{enumerate}

\begin{figure}[t!]
\centering
\includegraphics[width=\linewidth]{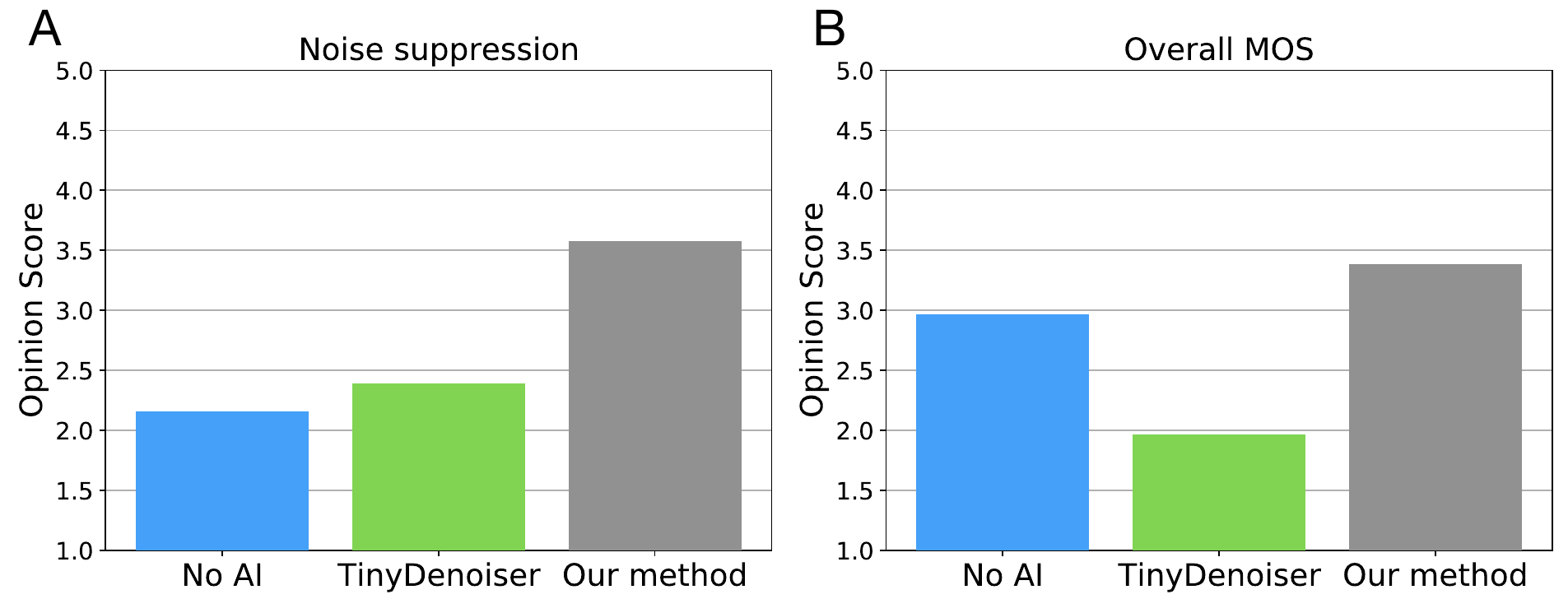}
\vskip -0.11in
\caption{Subjective in-the-wild evaluations. \textmd{ (a) Mean opinion score for the noise suppression quality reported for the three modes of operation, and (b) overall reported mean opinion score on the speech quality. }}
\label{fig:mosresult}
\vskip -0.05in
\end{figure}

{\bf Results.} As shown in Fig.~\ref{fig:mosresult}, our system significantly reduces background noise, as demonstrated by the increase in the mean opinion score for the noise suppression task from 2.15 in the “no AI” setting to 3.57. Additionally, our NeuralAids framework improved the overall mean opinion score (MOS) from 2.96 to 3.38.

Notably, while TinyDenoiser improves the average noise suppression score from 2.15 in the "no AI" setting to 2.38, it reduces the overall mean opinion score (MOS) from 2.96 to 1.96. This is because, although TinyDenoiser effectively reduces noise, it also significantly degrades speech quality by introducing distortions and choppiness. As a result, understanding the extracted speech becomes more difficult. This outcome aligns with our quantitative benchmarking results from~\xref{sec:benchmarks}. The likely reason   is that TinyDenoiser, designed to minimize computational complexity, incorporates only a minimal number of  components, making it challenging to both suppress noise and preserve speech quality effectively.

An important observation was that the system effectively adapts to sudden and rapid changes in the speaker’s position caused by the wearer rotating their head. Many participants in the study frequently turned their heads to look at different sound sources and objects in their surroundings. To assess this quantitatively, we evaluated the performance of a fine-tuned model across angular velocities ranging from 10 deg/s to 90 deg/s using our synthetic test set.  As shown in Table.~\ref{tab:speed}, the fine-tuned system demonstrates slightly better performance than the non-fine-tuned version and exhibits robustness to angular velocities of up to 90~deg/s. Notably, the non-fine-tuned version still performs well on samples with moving speakers despite not being trained on them. This could be due to the fact that this network operates on left and right audio channels independently and it does not use time-difference features which vary drastically with motion.

%% file: related-1.tex
\section{Related Work}

{\bf AI-based  enhanced hearing systems.} Recent  systems, such as Clearbuds~\cite{chatterjee2022clearbuds},  enhance  speech of the user wearing the earbuds. However, the processing is not on-device but instead occurs on a smartphone. Further, the target application is telephony, with delay constraints of 100–200 ms, which is an order of magnitude higher than our system.

Enhanced hearing systems like Semantic Hearing~\cite{semantichearing} allow users to pick and choose which classes of sounds they want to hear (e.g., car honks). This relies on a wired headset, with processing performed on an attached smartphone that has significantly higher processing capabilities. Similarly, prior work on target speech hearing~\cite{lookoncetohear} and sound bubbles~\cite{soundbubble} enable users to hear target speakers based on user-selected characteristics or distance. These require multiple microphones distributed across a wired over-the-head headset,  with processing performed on an external platform like an Orange Pi or a Raspberry Pi. 

Unlike all these prior works, our system operates on fully wireless hearables and addresses the fundamental question of whether one can design wireless hearables with programmable low-power AI accelerators and if deep learning models for speech processing can be designed to run on these wireless hearables in real time while meeting the  size, power, and compute constraints of these wireless hearable platforms.

\vskip 0.05in\noindent{\bf Commercial hearables.} Conventional hearables use statistical signal processing for speech enhancement. However,  deep neural networks have been demonstrated to achieve superior source separation, outperforming traditional methods by up to 9 dB~\cite{luo2019conv,chatterjee2022clearbuds}. Companies like Google~\cite{GooglePixelBudsAI} have begun integrating AI accelerators, such as the Tensor A1 chip in Pixel earbuds, though specific technical details of applications remain undisclosed. Similarly, in late 2024, Phonak~\cite{Phonak} introduced AI-powered hearing aids.  These commercial solutions remain proprietary, often provide limited technical detail, making fair and in-depth comparisons challenging. This highlights the need for academic exploration into the design space of  programmable, low-power speech AI for hearables and a better understanding of the challenges of running streaming deep learning models on hearables—a gap our work   addresses.

\begin{table}[t!]
  \caption{Motion results. \textmd{{Enhancement quality (SISDRi) with and without motion fine-tuning (FT). Results are shown for different ranges of angular speeds (deg/s).}}}
  \label{tab:motion}
  \vskip -0.1in
  \centering
\setlength{\tabcolsep}{2.4pt}
 
  \begin{tabular}{ l c c c c c c}
    \toprule
    \multicolumn{1}{l}{{Speed}} &
    \multicolumn{1}{c}{(-90,-60)} & 
    \multicolumn{1}{c}{{(-60,-30)}} &
    \multicolumn{1}{c}{{(-30,0)}} &
    \multicolumn{1}{c}{{(0,30)}} &
    \multicolumn{1}{c}{{(30,60)}} &
     \multicolumn{1}{c}{{(60,90)}} \\
    \midrule
    FT & 10.71 & 10.28 & 10.46 & 10.44 & 10.24 & 10.49  \\ 
     W/o FT & 10.42 & 9.95& 10.19 & 10.18 & 9.99 & 10.20 \\
    \bottomrule
  \end{tabular}
  \label{tab:speed}
   \vskip -0.2in
\end{table}

\vskip 0.05in\noindent{\bf Hearable platforms.} Existing   platforms have  been designed to enable earable research~\cite{10.1145/3636534.3649366,10.1145/3560905.3568084,10.1145/3447993.3448624,10.1145/3372224.3419197}, but none support on-device speech AI acceleration. The  Nokia Lab eSense platform~\cite{esense} introduced sensor-integrated earbuds, enabling data collection and application development for physiological sensing and tracking. OpenEarable~\cite{10.1145/3544793.3563415}, OpenEarable 2.0~\cite{10.1145/3712069}, and ClearBuds~\cite{chatterjee2022clearbuds} further advanced open-source earable platforms. However, the OpenEarable series lacks AI acceleration, relies  on a DSP chip  and focuses on physiological sensing~\cite{10.1145/3458864.3467680}. ClearBuds, meanwhile, uses a compute-limited microcontroller, making real-time on-device deep learning challenging. Similarly, the OpenMHA platform~\cite{openmha} lacks a hearable or hearing aid form factor and does not include any AI accelerator.

OmniBuds~\cite{montanari2024omnibudssensoryearableplatform}  expanded eSense’s bio-sensing capabilities and introduced on-board machine learning for physiological signal processing and classification tasks. However, it has not demonstrated speech AI, which is computationally more demanding. Additionally, it remains closed-source,  and even the specific chip used for on-device ML has not been disclosed -- the authors contacted the OmniBuds team, who could not provide this information. Speech AI requires significantly more computational power than classification tasks due to its high sampling rate, extensive I/O requirements, real-time constraints, and the need for causal processing within sub-10 ms latency. Prior to our work, achieving this on low-power hearables was considered  challenging.


\vskip 0.05in\noindent {\bf Low-latency speech processing.} Applications have different latency constraints in that they either can wait to process an entire audio file or require much quicker responses. In  augmented hearing, minimizing latency between input and processed output is crucial. However, this can degrade performance due to less available information for predicting the output~\cite{wang2022stft}.   Prior work has proposed  architectures for low-latency speech tasks~\cite{percepnet,speakerbeamss,deepfilternet2,lookoncetohear,waveformer}, and some speech challenges focus on these low-latency models~\cite{clarityChallenge}.  But, these are  designed for devices with significantly higher clock frequencies, power budgets, and memory footprints than our target hardware and none of these neural networks meet the computational constraints of wireless hearables. 

\vskip 0.05in\noindent{\bf Deep learning with  computational contraints.} Common methods include quantization~\cite{polino2018modelcompressiondistillationquantization}, pruning, and knowledge distillation~\cite{2stepdistillation}. In the audio domain, these techniques have been applied to tasks such as keyword spotting~\cite{kws}, speaker verification~\cite{tinysv}, and sound event detection~\cite{bose}. Prior work has also explored these techniques for speech enhancement and denoising~\cite{ cohen23_interspeech, TRunet, 10248154, 2stepdistillation, 8489456}. However, these models are not designed to operate in real-time on hearable hardware. Moreover, they do not process time and frequency components as individual sequences—an essential feature of state-of-the-art enhancement models. \cite{fqss} proposes a quantized network, but it  is neither causal nor real-time. 

The closest works to ours are TinyLSTM~\cite{tinylstms} and TinyDenoiser~\cite{tinydenoiser}, recurrent neural network (RNN)-based methods for speech enhancement. {TinyDenoiser has been shown to run in real-time on GAP9.} However, both models {have} an algorithmic latency of 25~ms. Furthermore, as shown in~\xref{sec:quantresults}, when modified for lower latency, performance degrades significantly. In contrast, we present the first dual-path speech enhancement network capable of real-time operation on low-power AI accelerators. We also explore mixed-precision quantization and quantization-aware training, evaluating various trade-offs concerning run-time performance on the target accelerator platform.

%% file: discuss-1.tex
\section{Limitations and Discussion}
The power consumption figures reported in this paper assume that the AI accelerator is continuously active, processing audio at all times. However, in real-world usage, energy consumption and battery life depend on how often the AI accelerator is actually in use. A more efficient design could activate the AI accelerator only when significant background noise is detected, allowing it to operate in a duty-cycled manner and significantly reducing average energy consumption. The effectiveness of this approach would depend on how frequently the wearer encounters noisy environments throughout the day, and the noise threshold, which is user dependent. Exploring such an adaptive system would be an interesting direction for future research.

Our goal was to demonstrate that effective speech AI models can run on low-power wireless hearables. We address this fundamental challenge using the GAP9 accelerator. However, other low-power AI accelerators are becoming increasingly available including the Analog Devices MAX78000 and MAX78002, Arm Ethos U-55 and U-85, and Kendryte K210, K230, and K510. Investigating their capabilities for running speech AI models would be  important  for future   research. 


Recent research has explored techniques with wired headsets  like semantic hearing~\cite{semantichearing}, sound bubbles~\cite{soundbubble}, and target speech hearing~\cite{lookoncetohear}. While we do not demonstrate the feasibility of implementing these capabilities on wireless hearables, it is worth noting that some of these systems~\cite{soundbubble,lookoncetohear} rely on the dual-path model architecture that we optimize in this paper. Further, recent  models like TF-MLPNet~\cite{tfmlpnet} achieve promising results for speech separation using  tiny quantized models. This suggests a potential path for expanding to these  capabilities in the near future.

Our hardware currently utilizes passive noise cancellation, where the earbuds physically block external sounds. However, we can enhance isolation by incorporating active noise cancellation (ANC). This would involve implementing ANC signal processing algorithms, which must meet strict delay requirements. The GAP9 hardware supports ANC algorithms, making  integration feasible for future  iterations.

Future work could also integrate additional sensors, such as PPG and temperature sensors, to support non-speech applications like  physiological sensing~\cite{montanari2024omnibudssensoryearableplatform}. Although our hardware has multiple microphones per device, our neural networks use only one. Exploring multi-microphone processing on a single hearable and across both ears could enhance performance and warrants further investigation.  

Since the inclusion criteria in our user study does not focus on hearing loss, future research is needed to incorporate signal-processing-based personalization algorithms tailored to individuals with hearing loss. This would involve adapting the system based on medical hearing loss prescriptions and systematically evaluating its effectiveness across different levels of hearing impairment.

Our paper focuses on behind-the-ear hearing aid form factor devices. The next step in this research would be integrating these speech AI models into earbud-form-factor devices, which is likely feasible due to the compact size of the target low-power AI accelerator.

Finally, the presence of a real-time AI accelerator with sufficient I/O support not only enables enhanced hearing capabilities but also paves the way for a broad spectrum of augmented intelligence applications. These may include  situational awareness, personal voice assistants, and cognitive support tools that extend beyond hearing enhancement. Thus, platforms such as ours that have embedded AI accelerators can enable  broader research and development in edge-AI-powered intelligence augmentation.

%% file: conclude-2.tex
\section{Conclusion}
The emergence of low-power programmable  AI accelerators presents an opportunity to bridge state-of-the-art speech AI with wireless earbuds and hearing aids. In this paper, we introduce NeuralAids, a fully on-device speech AI system for real-time speech enhancement and denoising on wireless hearables.  Our real-world evaluations  highlight the feasibility of deploying advanced speech AI models   on low-power wireless hearables, paving the way for next-gen intelligent audio devices that achieve on-device enhanced hearing.


\begin{acks}
The researchers are partly supported
by the Moore Inventor Fellow award \#10617, UW WE-REACH grant, Thomas J. Cable Endowed Professorship, and  a UW CoMotion innovation gap fund. This work was facilitated through the use of computational, storage, and networking infrastructure provided by the UW HYAK Consortium.
\end{acks}